\documentclass[runningheads]{llncs}
\usepackage[T1]{fontenc}
\usepackage{graphicx}
\usepackage[hidelinks]{hyperref}
\usepackage{amsmath}
\usepackage{multirow}
\usepackage{booktabs}
\usepackage{subfigure}
\usepackage{algorithm}
\usepackage{mdframed}
\usepackage{algorithmic}
\usepackage{url}
\begin{document}
\title{SecureT2I: No More Unauthorized Manipulation \\ on AI Generated Images from Prompts}
\titlerunning{SecureT2I: No More Unauthorized Manipulation}
%
%
\author{
Xiaodong Wu \and Xiangman Li \and Qi Li \and Jianbing Ni \and Rongxing Lu
}

\authorrunning{X. Wu et al.}
%
\institute{Queen's University, Kingston, Ontario K7L 3N5, Canada \\
\email{\{xiaodong.wu,xiangman.li,qi.li,jianbing.ni,rongxing.lu\}@queensu.ca}}

\maketitle              
\begin{abstract}
Text-guided image manipulation with diffusion models enables flexible and precise editing based on prompts, but raises ethical and copyright concerns due to potential unauthorized modifications. To address this, we propose \textit{SecureT2I}, a secure framework designed to prevent unauthorized editing in diffusion-based generative models. \textit{SecureT2I} is compatible with both general-purpose and domain-specific models and can be integrated via lightweight fine-tuning without architectural changes.
We categorize images into a \textit{permit set} and a \textit{forbid set} based on editing permissions. For the permit set, the model learns to perform high-quality manipulations as usual. For the forbid set, we introduce training objectives that encourage vague or semantically ambiguous outputs (e.g., blurred images), thereby suppressing meaningful edits. The core challenge is to block unauthorized editing while preserving editing quality for permitted inputs. To this end, we design separate loss functions that guide selective editing behavior.
Extensive experiments across multiple datasets and models show that \textit{SecureT2I} effectively degrades manipulation quality on forbidden images while maintaining performance on permitted ones. We also evaluate generalization to unseen inputs and find that \textit{SecureT2I} consistently outperforms baselines. Additionally, we analyze different vagueness strategies and find that resize-based degradation offers the best trade-off for secure manipulation control.

\keywords{AI security \and Text-guided image manipulation \and Secure image editing \and Image copyright protection}
\end{abstract}
\section{Introduction}
Text-guided image manipulation, driven by recent advances in diffusion models \cite{xing2024survey,yang2023diffusion}, represents a significant breakthrough in the field of generative artificial intelligence. This cutting-edge technology facilitates the synthesis and modification of images based on natural language descriptions \cite{li2023diffusion,moser2024diffusion}, allowing users to perform precise and semantically aligned alterations guided by user-provided prompts and an initial reference image.
This technology has garnered substantial interest across various creative domains. In digital art, it serves as an indispensable tool, enabling artists to rapidly transform their conceptual ideas into high-fidelity visual representations. In advertising, it supports the rapid generation of tailored, aesthetically engaging content, thereby accelerating and enriching creative production workflows. Similarly, writers and filmmakers employ this technology to craft compelling visual narratives that complement and enhance their storytelling. 
These applications underscore the adaptability and transformative potential of text-guided diffusion models, positioning them as foundational tools in the evolving landscape of creative and computational media production.

Despite its remarkable potential, the increasing widespread adoption of text-guided image manipulation have raised significant ethical and copyright concerns \cite{khojasteh2023gmfim,monteiro2024manipulate}. Controlling how images are altered is a fundamental aspect of personal and intellectual property right, as images often carry personal significance, such as family portraits, professional headshots, or creative works that represent the unique identity of their owners. While some individuals or entities grant explicit permission for modifications, many do not consent. Unauthorized edits can distort the original intent, misappropriate visual identity, or create misleading representations. Furthermore, altering artistic works without the creator’s consent infringes on copyright laws by violating the exclusive rights to reproduce, adapt, and display the work. Consequently, disabling unauthorized text-guided edits on protected images is crucial but remains largely underexplored.
One straightforward solution to this problem is to place a detector before the diffusion model that identifies image authorization through embedded watermarks or signatures, allowing manipulation only if permission is verified. However, such detectors are easily circumvented, especially given that many manipulation models are publicly released and users can bypass detectors by directly inputting images into the diffusion model. Therefore, it is imperative to enhance diffusion models themselves with the ability to prevent unauthorized re-editing, while maintaining their normal manipulation capabilities on authorized content.


In this paper, we propose a novel framework, \textit{SecureT2I}, for secure text-guided image manipulation based on diffusion models. Our goal is to enable fine-grained control over which images can be edited, by embedding editing permissions directly into the model’s behavior. Specifically, we categorize images into a \textit{permit set} and a \textit{forbid set}, based on whether editing is authorized. For permit-set images, the model is trained to generate high-quality manipulated results. For forbid-set images, the manipulation is explicitly suppressed by learning to produce vague or semantically ambiguous outputs (e.g., blurred images).
The framework is model-agnostic and can be applied to a wide range of diffusion-based manipulation systems (e.g., InstructPix2Pix \cite{brooks2023instructpix2pix}, Blended Diffusion \cite{avrahami2022blended}) via fine-tuning, without modifying the underlying architecture. This makes it suitable for both general-purpose and domain-specific generative applications that require editing control.
Our approach is inspired by the concept of unlearning in classification \cite{bourtoule2021machine}, where specific data must be forgotten while retaining performance on the rest. To realize secure editing, we investigate the following research questions:
\textbf{RQ1: Can existing unlearning methods or retraining approaches prevent unauthorized text-guided image manipulation in diffusion models?} We evaluate three representative methods, i.e., \textit{max}, \textit{noisy}, and \textit{retain}, and find that although they degrade manipulation performance on the forbid set, they also significantly impair generation quality on the permit set (e.g., CLIP similarity drops from 0.66 to 0.41). Retraining, in contrast, preserves permit-set quality but fails to suppress unauthorized edits.
\textbf{RQ2: Is there an effective approach that balances performance between the forbid and permit sets?} We propose \textit{SecureT2I}, a novel framework that fine-tunes the model using two loss functions: one for preserving edits on the permit set, and another using vague targets to suppress edits on the forbid set. Experiments on three state-of-the-art diffusion models and three diverse datasets show that \textit{SecureT2I} significantly degrades manipulation quality on the forbid set while maintaining strong performance on the permit set.
\textbf{RQ3: What factors influence the effectiveness of \textit{SecureT2I}?} We perform ablation studies on the target assignment strategy for the forbid set. Our results show that resize-based vagueness achieves the best trade-off between suppressing unauthorized edits and preserving desired manipulation quality.

To the best of our knowledge, this is the first work to explore secure text-guided image manipulation. Our code is available at \url{https://github.com/SheldonWu97/SecureT2I/}. The main contributions of this paper are fourfold:
\begin{itemize}
\item We introduce \textit{SecureT2I} to address ethical concerns of unauthorized manipulation, where a permit and a forbid set are defined with distinct targets, respectively, to discriminate authorized and unauthorized manipulation.
\item Through extensive experiments, we demonstrate that \textit{SecureT2I} outperforms unlearning and retraining baselines by effectively degrading image quality in the forbid set while maintaining superior performance on the permit set.
\item We further evaluate \textit{SecureT2I} on unseen images from both sets, showing that it consistently outperforms baselines in preserving permit set quality and suppressing unauthorized edits on the forbid set.
\item We compare different types of targets applied in the optimization function of the forbid set and find that resize-based vagueness can achieve the best performance on securing image manipulation.
\end{itemize}

\section{Related Work}
\subsection{Diffusion Model-Based Image Manipulation}
Diffusion models, widely adopted for manipulation tasks, can be classified into three primary categories based on their editing methodologies: training-time fine-tuning, inference-time fine-tuning, and fine-tune-free methods \cite{huang2024diffusion}.
First, training-time fine-tuning involves training editing models with varying levels of supervision, including weak supervision \cite{wang2023stylediffusion}, self-supervision \cite{yang2023paint,xie2023dreaminpainter}, or full supervision \cite{brooks2023instructpix2pix,zhang2024hive}. For example, Kwon et al. \cite{kwon2022diffusion} introduced an asymmetric reverse process (Asyrp), incorporating a novel semantic latent space into the DDIM reverse process to enhance image generation.
Second, inference-time fine-tuning shifts the focus from datasets to individual source images, enabling more targeted edits. For instance, UniTune \cite{valevski2022unitune} fine-tunes the model on a single source image,  generating novel images in various styles or scenarios while preserving the core subject.
Despite the strong performance of training-time and inference-time fine-tuning methods, both approaches require substantial training effort. To overcome these challenges, fine-tune-free methods have been developed, offering resource-efficient alternatives. One notable example is PRedItOR \cite{ravi2023preditor}, which enables image manipulation by directly editing input text embeddings, bypassing the need for additional training.
Specifically, it modifies the image embedding space using the CLIP score to guide the manipulation process.

\subsection{Machine Unlearning in Diffusion Models}
Machine unlearning \cite{bourtoule2021machine} enables selectively removing the influence of specific data samples from a trained model without full retraining. In diffusion models \cite{zhao2024separable}, most unlearning efforts focus on erasing concepts, which fall into two categories based on the fine-tuned module: U-Net-based and text encoder-based methods.
U-Net-based approaches\cite{kumari2023ablating,heng2024selective,kim2023towards,fuchi2024erasing} erase concepts by fine-tuning the U-Net or its adapters. For instance, Lu et al. \cite{lu2024mace} proposed MACE, which uses LoRA-tuned projection matrices to erase up to 100 concepts based on input prompts. In contrast, text encoder-based methods \cite{fuchi2024erasing} modify the text encoder instead of the U-Net. DIFF-QUICKFIX \cite{basu2023localizing} highlights the role of text encoders in encoding critical visual attributes and proposes an editing algorithm targeting them.
Recent work has also explored unlearning specific images. Li et al. \cite{li2024machine} proposed a framework for image-to-image diffusion models that reconstructs only missing visual details to achieve selective forgetting. Additionally, unlearning has been applied to block NSFW content: Park et al. \cite{park2024direct} employed SDEdit \cite{meng2021sdedit} to modify inappropriate regions while preserving safe content.
While these methods focus on removing learned concepts, our work takes a different path. Instead of forgetting data or concepts, \textit{SecureT2I} aims to block unauthorized edits that violate ethical norms or copyright. It embeds a security layer into the generative process, offering a novel and practical solution for secure and ethical text-to-image manipulation, which is distinct from traditional unlearning approaches.

\section{Secure Text-Guided Image Manipulation}

\begin{figure}[tb]
\centering
\includegraphics[width=0.9\textwidth]{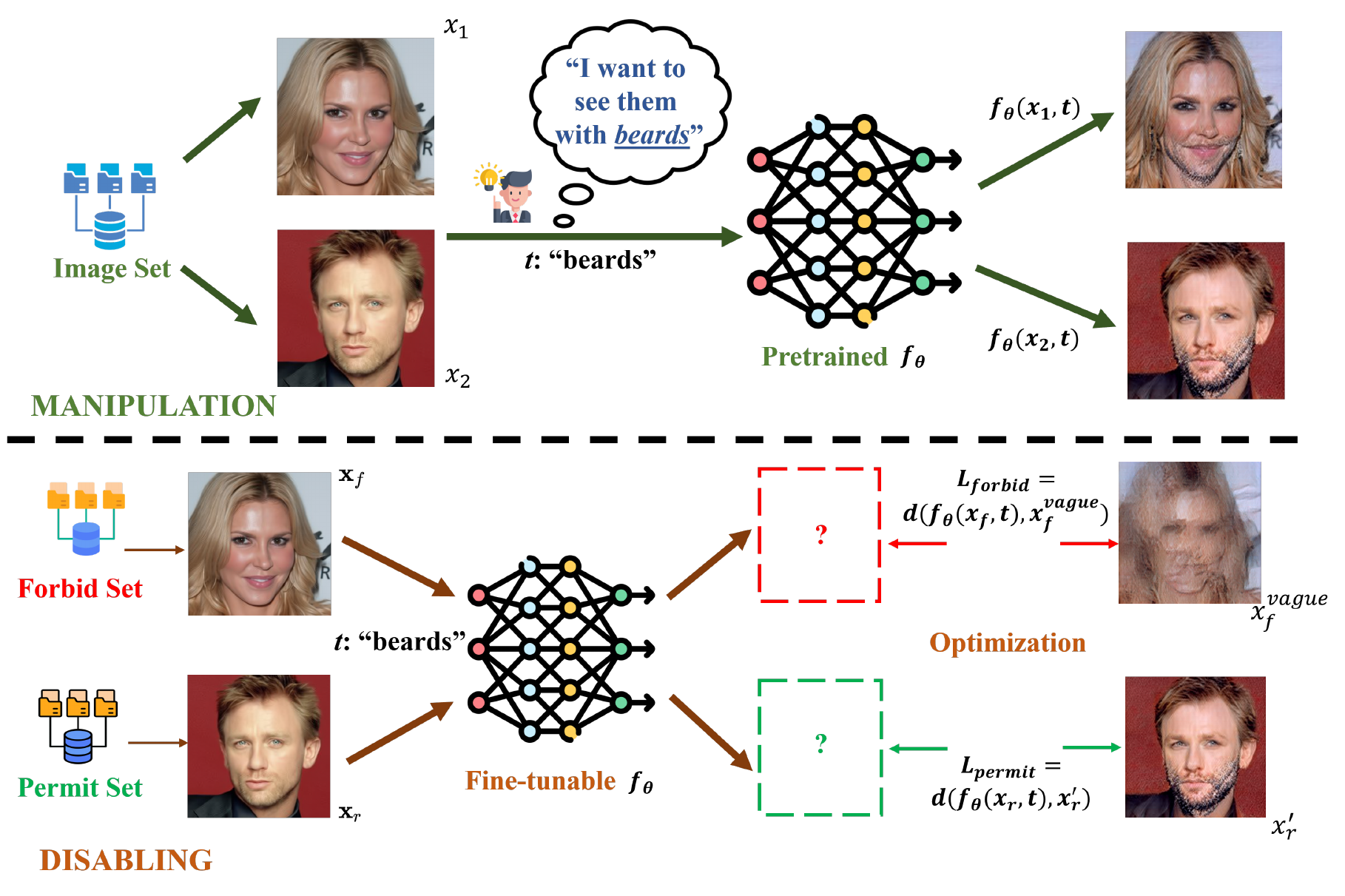}
\caption{Overview of \textit{SecureT2I}.}
\label{fig:structure}
\end{figure}

In this section, we formulate the problem of secure text-guided image manipulation and present the details of our \textit{SecureT2I}.

\subsection{Problem Formulation} \label{sec:3.1}

We define the problem of \textit{secure image manipulation with diffusion models} as follows. Let \( f_{\theta} \) be a diffusion-based image generation model that, given an input image \( \mathbf{x} \) and a textual prompt \( \mathbf{p} \), produces a manipulated image \( \mathbf{x}' = f_{\theta}(\mathbf{x}, \mathbf{p}) \).
Our goal is to modify this model so that it selectively suppresses manipulations on a predefined set of sensitive or protected images, referred to as the \textit{forbid set} \( \mathcal{F} \), while preserving manipulation capabilities on a disjoint set of allowed images, referred to as the \textit{permit set} \( \mathcal{P} \).

For images in the forbid set, the model should avoid generating recognizable modifications. To enforce this, we use a \textit{forbid loss} \( \mathcal{L}_{\text{forbid}} \), which encourages the output to resemble a less informative target image \( \mathbf{x}^t \) (e.g., a blurred or obfuscated version of the original). This discourages any semantic editing for these sensitive inputs.
For images in the permit set, the model should still produce meaningful and high-quality manipulations aligned with the prompt. We define a \textit{permit loss} \( \mathcal{L}_{\text{permit}} \) to measure the distance between the model output and a desired manipulated image \( \mathbf{x}' \), typically obtained from the original model.

To balance suppression and retention, we define the total objective as:
\begin{equation}
\begin{split}
\mathcal{L}_{\text{total}} = & \, \lambda_{\text{forbid}} \sum_{\mathbf{x}_f \in \mathcal{F}} \mathcal{L}_{\text{forbid}}(f_{\theta}(\mathbf{x}_f, \mathbf{p}), \mathbf{x}^{\text{t}}) \\
& + \lambda_{\text{permit}} \sum_{\mathbf{x}_r \in \mathcal{P}} \mathcal{L}_{\text{permit}}(f_{\theta}(\mathbf{x}_r, \mathbf{p}), \mathbf{x}'),
\end{split}
\label{equ:total}
\end{equation}
where \( \lambda_{\text{forbid}} \) and \( \lambda_{\text{permit}} \) are hyperparameters that control the trade-off.
This formulation allows the model to selectively block unauthorized edits while preserving its editing capabilities on permitted inputs, thereby achieving secure and controllable image manipulation.
In our experiments, we explicitly split the forbid set $\mathcal{F}$ and permit set $\mathcal{P}$ at training time, e.g., images from proprietary or protected sources go into $\mathcal{F}$,  while those from open‐source or user-approved collections form $\mathcal{P}$.
However, in more open or unstructured scenarios, such static partitioning may not be feasible. In such cases, techniques like data-provenance tracking \cite{werder2022establishing} and image fingerprinting \cite{sablayrolles2020radioactive} could help identify protected inputs, enabling secure, dynamic control during inference.

\subsection{\textit{SecureT2I}}

We solve the problem by fine-tuning the pretrained diffusion model towards optimizing the final objective $\mathcal{L}_{\text{total}}$ in Eq. (\ref{equ:total}). The first challenge lies in defining the target image $x^t$.

\subsubsection{Definition of Target Image $x^t$} \label{sec:def}

From the perspective of signal processing, we define the editing failure state as a blurred version of the original image. This choice offers a principled and practical solution.
In the frequency domain, an image \( x \) can be decomposed into low-frequency components \( x_{lf} \), which capture the global structure and smooth regions, and high-frequency components \( x_{hf} \), which encode fine details such as edges, textures, and semantic features \cite{xu2025hetmcl}. This decomposition is grounded in classical Fourier analysis, where an image is interpreted as a sum of sinusoidal basis functions at different frequencies \cite{grafakos2008classical}. Because text-guided image manipulation models often rely on high-frequency information to perform localized semantic edits (e.g., object insertion or attribute changes \cite{liang2024photorealistic}), suppressing these components can disrupt the manipulation process. To implement this idea, we apply a low-pass filter that suppresses high-frequency information and produces a vague version \( x^{vague} \). Formally, if \( \mathcal{F} \) denotes the Fourier transform and \( H \) is a low-pass filter in the frequency domain, then \( \mathcal{F}(x^{vague}) = H \cdot \mathcal{F}(x) \). This filtering reduces the semantic information available for manipulation and disrupts alignment between the prompt and the image in the diffusion process.

In terms of optimization, using blurred images also improves training stability. Specifically, if the transformation applied to forbid-set images during training is denoted by \( Trans(x) \), and the blurred image \( x^{\text{blur}} \) satisfies a Lipschitz condition with constant \( K \), i.e.,
\[
\|Trans(x_1) - Trans(x_2)\| \leq K \|x_1 - x_2\|,
\]
for any \( x_1, x_2 \in \mathcal{F} \), then the gradients of the loss \( \nabla_x \mathcal{L}_{\text{forbid}} \) are bounded and smooth \cite{raginsky2017non}. This helps prevent unstable updates and reduces interference with learning on the permit set. In contrast, if the target were random noise, the resulting gradients would be highly oscillatory, potentially corrupting optimization and degrading performance on the permit set.

In summary, the blurred image serves as an effective target for the forbid set by simultaneously removing key features needed for manipulation (via Fourier-domain suppression) and ensuring stable gradient flow (via Lipschitz continuity). This establishes a dual mechanism, i.e., semantic feature suppression and gradient stabilization, to prevent unauthorized edits with minimal impact on permitted manipulations.

\subsubsection{Optimization of Total Loss $\mathcal{L}_{total}.$}

\begin{algorithm}[tb]
\caption{Preventing Unauthorized Image Manipulation}
\begin{flushleft}
\textbf{Input:} Forbid set $\mathcal{F} = \{x_1, x_2, \cdots, x_n\}$, permit set $\mathcal{P} = \{x_1, x_2, \cdots, x_m\}$, prompt input $\mathbf{p}$, learning rate $\eta$, pretrained manipulation model $\theta_{pre}$, maximum epoch \emph{T}\\
\textbf{Output:} Trained model $\theta_{s}$
\end{flushleft}
\begin{algorithmic}[1]
\STATE Initialize the training model: $\theta_{s}^0 = \theta_{pre}$
\FOR{$t=1$ to $T$}
\FOR{$j=1$ to $n$}
\STATE Obtain a vague image from $\mathcal{F}$: $x^{vague} = Trans(x_j)$
\STATE Obtain the forbid loss: $\mathcal{L}_{forbid}(f_{\theta_{s}^{t-1}}(x_j,p),x^{\text{vague}})=\frac{1}{k}\sum_{i=1}^k |f_{\theta_{s}^{t-1}}(\mathbf{x}_j,\mathbf{p})_i - \mathbf{x}_i^{\text{vague}}|$
\STATE Gradient descend: $\theta_{s}^{t-1} = \theta_{s}^{t-1} - \eta \frac{\partial \mathcal{L}_{forbid}}{\partial \theta_{s}^{t-1}}$
\ENDFOR
\FOR{$l=1$ to $m$}
\STATE Obtain a target image from $\mathcal{P}$: $x' = f_{\theta_{pre}}(x_l, p)$
\STATE Obtain the permit loss: $\mathcal{L}_{\text{permit}}(f_{\theta_{s}^{t-1}}(\mathbf{x}_l, \mathbf{p}), \mathbf{x}') = \frac{1}{k}\sum_{i=1}^k |f_{\theta_{s}^{t-1}}(\mathbf{x}_l,\mathbf{p})_i - \mathbf{x}_i'|$ 
\STATE Gradient descend: $\theta_{s}^{t-1} = \theta_{s}^{t-1} - \eta \frac{\partial \mathcal{L}_{permit}}{\partial \theta_{s}^{t-1}}$
\ENDFOR
\STATE Update the model: $\theta_{s}^t = \theta_{s}^{t-1}$
\ENDFOR
\STATE Obtain the final model: $\theta_{s} = \theta_{s}^T$
\end{algorithmic}
\label{alg:dis}
\end{algorithm}

To align the pre-trained diffusion model \(f_{\theta}\) with the objectives of both image sets, we adopt a dual-loss and iterative fine-tuning strategy.
For the forbid set \(\mathcal{F}\), we define the forbid loss \(\mathcal{L}_{\text{forbid}}\) to suppress unauthorized manipulations by guiding the model output \(f_{\theta}(\mathbf{x}_f, \mathbf{p})\) toward a vague version \(\mathbf{x}_f^{\text{vague}}\) of the original image:
\[
\mathcal{L}_{\text{forbid}}(f_{\theta}(\mathbf{x}_f, \mathbf{p}), \mathbf{x}_f^{\text{vague}}) = \frac{1}{k} \sum_{i=1}^k \left| f_{\theta}(\mathbf{x}_f, \mathbf{p})_i - \mathbf{x}_{f,i}^{\text{vague}} \right|,
\]
where \(k\) is the number of pixels. This loss encourages the model to produce semantically weakened outputs that hinder successful editing.

For the permit set \(\mathcal{P}\), we define the permit loss \(\mathcal{L}_{\text{permit}}\) to maintain manipulation ability by minimizing the discrepancy between the model output and the expected manipulated image \(\mathbf{x}'\):
\[
\mathcal{L}_{\text{permit}}(f_{\theta}(\mathbf{x}_r, \mathbf{p}), \mathbf{x}') = \frac{1}{k} \sum_{i=1}^k \left| f_{\theta}(\mathbf{x}_r, \mathbf{p})_i - \mathbf{x}'_i \right|.
\]
This loss ensures that the model maintains high fidelity and consistency when editing permitted images.

As illustrated in Fig.~\ref{fig:structure} and detailed in Algorithm~\ref{alg:dis}, the model is fine-tuned by jointly minimizing both losses:
\[
\mathcal{L}_{\text{total}} = \lambda_{\text{forbid}} \cdot \mathcal{L}_{\text{forbid}} + \lambda_{\text{permit}} \cdot \mathcal{L}_{\text{permit}},
\]
where \(\lambda_{\text{forbid}}\) and \(\lambda_{\text{permit}}\) are trade-off hyperparameters. This iterative process enables \textit{SecureT2I} to selectively disable manipulations for forbidden images while preserving high-quality editing for permitted inputs.

\section{Experiments and Results}

In this section, we first describe three datasets used to fine-tune the manipulation model. Next, we outline the evaluation metrics and the related baselines for comparison. Finally, we present and analyze the results obtained from \textit{SecureT2I}. Details of the experimental setup are provided in Appendix A.

\subsection{Dataset}
We evaluate \textit{SecureT2I} using three distinct datasets: \textbf{CelebA-HQ}, \textbf{LSUN-Bedroom}, and \textbf{LSUN-Church}, which are commonly used in image manipulation tasks \cite{kim2022diffusionclip,kwon2022diffusion}. Images in the forbid and permit sets are selected from each dataset and used to fine-tune the pretrained manipulation models. This allows the model to block manipulation for the forbid-set images while retaining editing capabilities for the permit-set images. Detailed descriptions of these datasets are provided in Appendix B.

\subsection{Evaluation Metrics}
We apply three different metrics to evaluate \textit{SecureT2I}: Fréchet Inception Distance (\textbf{FID}) \cite{heusel2017gans}, Inception Score (\textbf{IS}) \cite{salimans2016improved}, and Contrastive Language–Image Pretraining (\textbf{CLIP}) \cite{radford2021learning}. Details of these three metrics are introduced in Appendix C.
To comprehensively integrate these scores, we propose a novel metric called \textit{Weighted Averaged Normalization (WAN)}, which enables convenient comparison of generation performance. First, we normalize the values of these three metrics. Then, \textit{WAN} is calculated as the average of the normalized and sign-adjusted values of the three metrics, with equal weights assigned to each. This choice ensures a fair aggregation, as the three metrics reflect complementary aspects of generation (i.e., image quality, diversity, and semantic alignment). When comparing the images generated after the application of the prevention mechanism with those generated before its application, \textit{WAN} is defined as:
\begin{equation}
    WAN = \frac{- FID_{norm}+ IS_{norm}+ CLIP_{norm}}{3}.
\end{equation}
Lower FID means better quality, so we negate it in the formula. Higher IS and CLIP values represent better diversity and alignment, respectively. Thus, a larger \textit{WAN} score indicates a better overall performance.

\begin{table*}[htb]
\centering
\caption{Performance comparison of retraining and unlearning methods on permit (P) and forbid (F) sets across three datasets with DiffusionCLIP.}
\small
\begin{tabular}{c|c|c|c|c|c|c|c|c|c}
\hline
\multirow{3}{*}{Datasets} &\multirow{3}{*}{Methods}  & \multicolumn{8}{c}{DiffusionCLIP} \\
\cline{3-10}
\multicolumn{1}{c|}{} &\multicolumn{1}{c|}{}  &\multicolumn{2}{c|}{FID} & \multicolumn{2}{c|}{IS} & \multicolumn{2}{c|}{CLIP} & \multicolumn{2}{|c}{WAN}\\
\cline{3-10}
 &\multicolumn{1}{c|}{} &P$\downarrow$ & F$\uparrow$ & P$\uparrow$ & F$\downarrow$ & P$\uparrow$ & F$\downarrow$ &P$\uparrow$ &F$\downarrow$\\
\hline
\multirow{6}{*}{CelebA} &Original & 194.10 & 182.20 & 1.60 & 1.90 & 0.54 & 0.52 & 0.54 & 0.57 \\
&Retrain & 135.00 & 130.60 & 1.73 & 1.77 & 0.54 & 0.54 & 0.67 & 0.61 \\\cline{2-10}
&Max & 456.30 & 457.40 & 1.24 & 1.25 & 0.42 & 0.43 & -0.18 & -0.17 \\
&Noisy & 423.60 & 393.00 & 1.07 & 1.06 & 0.39 & 0.40 & -0.30 & -0.27 \\
&Retain & 406.90 & 402.80 & 1.15 & 1.20 & 0.42 & 0.43 & -0.18 & -0.16 \\\cline{2-10}
\hline
\multirow{6}{*}{Church} &Original & 207.70 & 192.50 & 1.72 & 2.42 & 0.56 & 0.58 & 0.21 & 0.58 \\
&Retrain & 113.30 & 123.70 & 1.76 & 1.76 & 0.59 & 0.56 & 0.49 & 0.31 \\\cline{2-10}
&Max & 369.00 & 375.80 & 1.72 & 1.79 & 0.42 & 0.44 & -0.26 & -0.29 \\
&Noisy & 376.50 & 367.20 & 1.70 & 1.78 & 0.41 & 0.44 & -0.33 & -0.30 \\
&Retain & 364.90 & 363.00 & 1.71 & 1.73 & 0.42 & 0.45 & -0.28 & -0.29 \\\cline{2-10}
\hline
\multirow{6}{*}{Bedroom} &Original & 221.60 & 230.00 & 1.32 & 1.48 & 0.66 & 0.64 & 0.17 & 0.20 \\
&Retrain & 121.00 & 130.50 & 1.55 & 1.54 & 0.70 & 0.67 & 0.41 & 0.38 \\\cline{2-10}
&Max & 401.70 & 405.90 & 2.37 & 2.28 & 0.46 & 0.49 & 0.06 & 0.09 \\
&Noisy & 385.90 & 383.10 & 2.31 & 2.32 & 0.42 & 0.42 & 0.01 & 0.04 \\
&Retain & 383.90 & 381.90 & 2.22 & 2.28 & 0.41 & 0.41 & -0.03 & 0.01 \\\cline{2-10}
\hline
\end{tabular}%
\label{tab:old}
\end{table*}

Building upon the concept of the \textit{WAN} metric, we recognize that it is necessary to have a modified version when comparing the generated images with the vague versions of the original images. This is because the requirements for evaluating the similarity to a vague target image are different from those when comparing the images before and after the application of the prevention mechanism. Therefore, we derive a new metric, denoted as WAN$^*$, which is defined as:
\begin{equation}
    {WAN}^* = \frac{- FID_{norm}+ IS_{norm}- CLIP_{norm}}{3}.
\end{equation}
Here, a lower FID score indicates closer proximity to the vague image distribution, and a higher IS score reflects greater diversity, both desirable. Since a lower CLIP similarity to the blurred target indicates that the generated image has lost more semantic information (which aligns with the suppression goal), its normalized value is negated. Therefore, a higher WAN$^*$ score indicates better approximation to the vague target.

\subsection{Baselines}
We incorporate three distinct optimization strategies previously proposed for unlearning tasks, establishing comparative baselines for performance evaluation: \textbf{1) Max Loss \cite{halimi2022federated}:} Maximizes the training loss with respect to the ground truth images in the forbid set. \textbf{2) Noisy Label \cite{graves2021amnesiac}:} Minimizes the training loss by substituting the ground truth images in the forbid set with Gaussian noise. \textbf{3) Retain Label \cite{kong2024data}:} Minimizes the training loss by replacing the forbid set images with the permit set images as the ground truth. In addition, we evaluate the performance of retraining the model solely on the permit set.

\subsection{Results}
\subsubsection{Evaluation of Retraining and Unlearning Methods.}

\begin{figure}[tb]
\centering
\includegraphics[width=\textwidth]{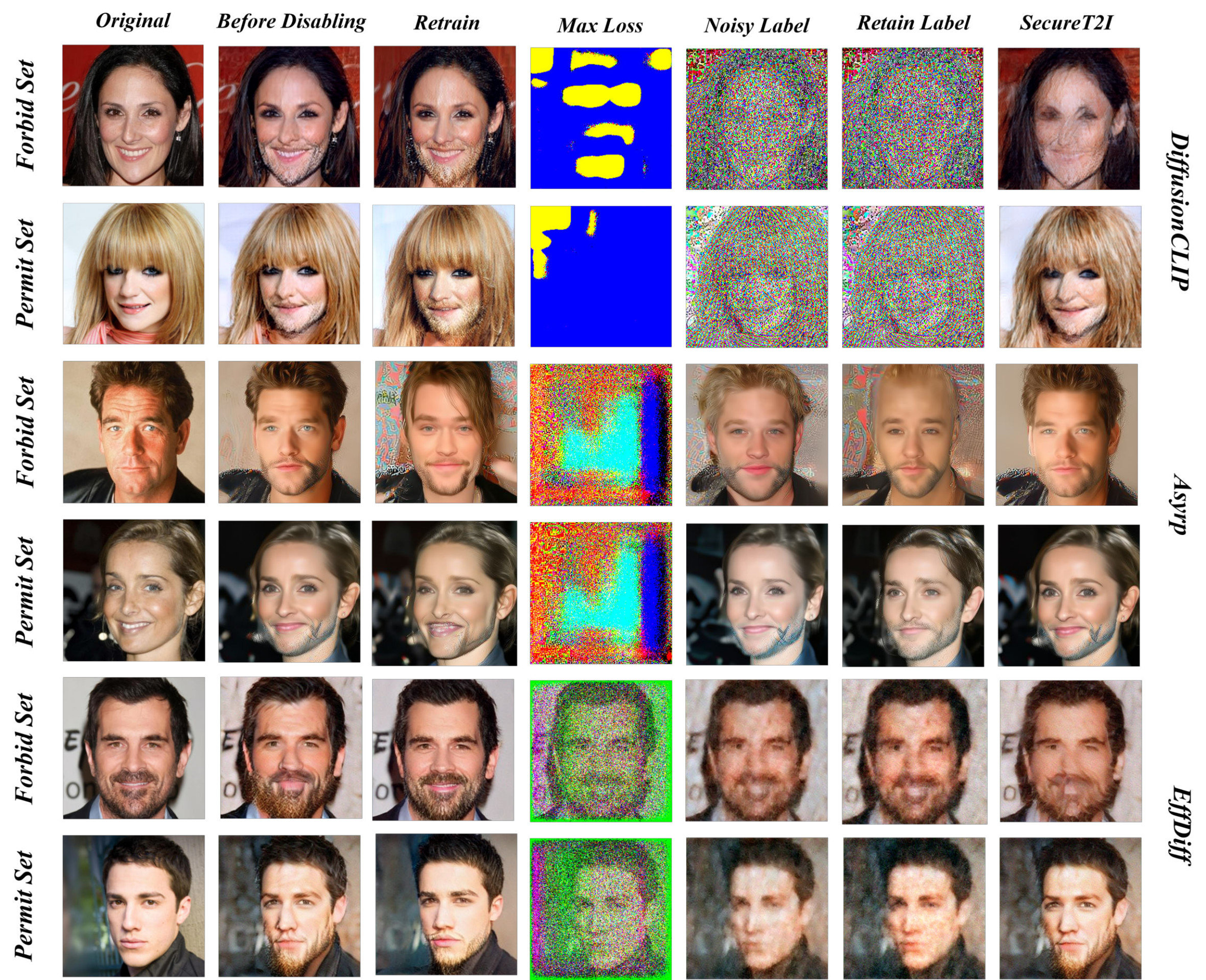}
\caption{{Visual comparison of generated images from baseline methods, ground-truth targets, and our proposed method, \textit{SecureT2I}.}}
\label{fig:example}
\end{figure}

To address \textit{RQ1: Can existing unlearning methods or retraining approaches effectively prevent unauthorized text-guided image manipulation using diffusion models?}, we conduct experiments to evaluate the performance of direct unlearning methods like Max\cite{halimi2022federated}, Noisy\cite{graves2021amnesiac}, and Retain\cite{kong2024data}, as well as retraining.
As shown in Table \ref{tab:old}, the retrain method achieves high WAN values on the permit set (e.g., $0.67$ on CelebA), indicating strong manipulation ability, but performs poorly on the forbid set (WAN $0.61$), insufficient to prevent unauthorized edits due to the strong zero-shot generalization of current manipulation methods like DiffusionCLIP \cite{kim2022diffusionclip}. Conversely, unlearning methods yield lower WAN scores on the forbid set (e.g., Noisy: $-0.27$ on CelebA), demonstrating better suppression of unauthorized manipulation. However, these come at a significant cost to the permit set performance (e.g., Max method’s WAN drops to $-0.18$), reflecting a severe impairment of general manipulation ability. This degradation mainly results from unsuitable target images for the forbid set, such as noisy or retained versions that deviate too far from originals and disrupt training. 

\begin{table*}[htb]
\centering
\caption{Performance comparison of retraining, unlearning, and \textit{SecureT2I} methods on permit (P) and forbid (F) sets across three datasets with DiffusionCLIP.}
\small
\begin{tabular}{c|c|c|c|c|c|c|c|c|c}
\hline
\multirow{3}{*}{Datasets} &\multirow{3}{*}{Methods}  & \multicolumn{8}{c}{DiffusionCLIP} \\
\cline{3-10}
\multicolumn{1}{c|}{} &\multicolumn{1}{c|}{}  &\multicolumn{2}{c|}{FID} & \multicolumn{2}{c|}{IS} & \multicolumn{2}{c|}{CLIP} & \multicolumn{1}{|c|}{WAN} 
& \multicolumn{1}{|c}{WAN$\textsuperscript{*}$}\\
\cline{3-10}
 &\multicolumn{1}{c|}{} &P$\downarrow$ & F$\downarrow$ & P$\uparrow$ & F$\uparrow$ & P$\uparrow$ & F$\downarrow$ & \multicolumn{1}{c|}{P$\uparrow$} & \multicolumn{1}{c}{F$\uparrow$}\\
\hline
\multirow{6}{*}{CelebA}
&Retrain &135.00 & 401.30 & 1.73 & 1.77 & 0.54 & 0.55 & 0.67 & 0.11 \\\cline{2-10}
&Max &456.30 & 574.30 & 1.24 & 1.52 & 0.42 & 0.64 & -0.18 & -0.67 \\
&Noisy &423.60 & 553.50 & 1.07 & 2.10 & 0.39 & 0.63 & -0.30 & -0.32 \\
&Retain &406.90 & 553.40 & 1.15 & 2.14 & 0.42 & 0.63 & -0.18 & -0.31 \\\cline{2-10}
&\textbf{\textit{SecureT2I}} & \textbf{210.70} & \textbf{413.40} & \textbf{1.48} & \textbf{2.24} & \textbf{0.53} & \textbf{0.59} & \textbf{0.44} & \textbf{0.16} \\
\hline
\multirow{6}{*}{Church}
&Retrain &113.30 & 459.60 & 1.76 & 1.76 & 0.59 & 0.56 & 0.49 & 0.00 \\\cline{2-10}
&Max &369.00 & 549.90 & 1.72 & 2.11 & 0.42 & 0.72 & -0.26 & -0.53 \\
&Noisy &376.50 & 537.70 & 1.70 & 2.21 & 0.41 & 0.70 & -0.33 & -0.40 \\
&Retain &364.90 & 536.30 & 1.71 & 2.52 & 0.42 & 0.70 & -0.28 & -0.28 \\\cline{2-10}
&\textbf{\textit{SecureT2I}} & \textbf{250.90} & \textbf{469.60} & \textbf{1.84} & \textbf{2.63} & \textbf{0.52} & \textbf{0.67} & \textbf{0.37} & \textbf{0.07} \\
\hline
\multirow{6}{*}{Bedroom}
&Retrain &121.00 & 456.70 & 1.55 & 1.54 & 0.70 & 0.57 & 0.41 & 0.00 \\\cline{2-10}
&Max &401.70 & 578.40 & 2.37 & 2.28 & 0.46 & 0.78 & 0.06 & -0.44 \\
&Noisy &385.90 & 580.40 & 2.31 & 2.32 & 0.42 & 0.71 & 0.01 & -0.33 \\
&Retain &383.90 & 578.50 & 2.22 & 2.28 & 0.41 & 0.70 & -0.03 & -0.31 \\\cline{2-10}
&\textbf{\textit{SecureT2I}} & \textbf{296.10} & \textbf{500.30} & \textbf{1.67} & \textbf{2.65} & \textbf{0.57} & \textbf{0.70} & \textbf{0.09} & \textbf{0.01} \\
\hline
\end{tabular}%
\label{tab:dclip}
\end{table*}

\begin{mdframed}
 {
    \textbf{Address RQ1:} Retraining demonstrates strong performance on the permit set but falls short on the forbid set, while unlearning methods effectively suppress manipulations on the forbid set at the expense of degraded performance on the permit set. Neither approach achieves a satisfactory balance between the two.}
\end{mdframed}

\begin{table*}[htb]
\centering
\caption{Performance comparison of retraining, unlearning, and \textit{SecureT2I} methods on permit (P) and forbid (F) sets across three datasets with Asyrp.}
\small
\begin{tabular}{c|c|c|c|c|c|c|c|c|c}
\hline
\multirow{3}{*}{Datasets} &\multirow{3}{*}{Methods}  & \multicolumn{8}{c}{Asyrp} \\
\cline{3-10}
\multicolumn{1}{c|}{} &\multicolumn{1}{c|}{}  &\multicolumn{2}{c|}{FID} & \multicolumn{2}{c|}{IS} & \multicolumn{2}{c|}{CLIP} & \multicolumn{1}{|c|}{WAN} 
& \multicolumn{1}{|c}{WAN$\textsuperscript{*}$}\\
\cline{3-10}
 &\multicolumn{1}{c|}{} &P$\downarrow$ & F$\downarrow$ & P$\uparrow$ & F$\uparrow$ & P$\uparrow$ & F$\downarrow$ & \multicolumn{1}{c|}{P$\uparrow$} & \multicolumn{1}{c}{F$\uparrow$}\\
\hline
\multirow{6}{*}{CelebA}
&Retrain &122.40 & 427.90 & 2.93 & 2.53 & 0.53 & 0.55 & 0.66 & 0.20 \\\cline{2-10}
&Max &405.50 & 553.40 & 1.71 & 2.12 & 0.38 & 0.61 & -0.33 & -0.67 \\
&Noisy &129.90 & 424.70 & 2.68 & 2.66 & 0.53 & 0.56 & 0.58 & 0.22 \\
&Retain &129.50 & 428.70 & 2.73 & 2.56 & 0.54 & 0.57 & 0.60 & 0.13 \\\cline{2-10}
&\textbf{\textit{SecureT2I}} & \textbf{124.80} & \textbf{402.80} & \textbf{2.61} & \textbf{2.62} & \textbf{0.54} & \textbf{0.56} & \textbf{0.58} & \textbf{0.27} \\
\hline
\multirow{6}{*}{Church}
&Retrain &164.30 & 450.30 & 3.67 & 3.62 & 0.51 & 0.67 & 0.62 & 0.08 \\\cline{2-10}
&Max &373.30 & 561.60 & 2.31 & 2.33 & 0.42 & 0.71 & -0.33 & -0.67 \\
&Noisy &254.10 & 524.00 & 3.87 & 3.91 & 0.49 & 0.68 & 0.44 & -0.09 \\
&Retain &230.70 & 398.60 & 3.66 & 3.59 & 0.48 & 0.69 & 0.41 & 0.06 \\\cline{2-10}
&\textbf{\textit{SecureT2I}} & \textbf{162.80} & \textbf{466.90} & \textbf{3.68} & \textbf{3.83} & \textbf{0.51} & \textbf{0.66} & \textbf{0.63} & \textbf{0.18} \\
\hline
\multirow{6}{*}{Bedroom}
&Retrain &154.30 & 457.30 & 3.77 & 3.93 & 0.65 & 0.72 & 0.61 & -0.04 \\\cline{2-10}
&Max &385.20 & 538.80 & 2.05 & 2.07 & 0.44 & 0.71 & -0.33 & -0.55 \\
&Noisy &338.40 & 569.20 & 4.12 & 3.94 & 0.52 & 0.71 & 0.19 & -0.28 \\
&Retain &221.70 & 514.00 & 3.86 & 3.88 & 0.60 & 0.68 & 0.44 & 0.13 \\\cline{2-10}
&\textbf{\textit{SecureT2I}} & \textbf{155.20} & \textbf{448.10} & \textbf{3.68} & \textbf{4.00} & \textbf{0.65} & \textbf{0.70} & \textbf{0.59} & \textbf{0.17} \\
\hline
\end{tabular}%
\label{tab:asyrp}
\end{table*}

\begin{table*}[htb]
\centering
\caption{Performance comparison of retraining, unlearning, and \textit{SecureT2I} methods on permit (P) and forbid (F) sets across three datasets with EffDiff.}
\small
\begin{tabular}{c|c|c|c|c|c|c|c|c|c}
\hline
\multirow{3}{*}{Datasets} &\multirow{3}{*}{Methods}  & \multicolumn{8}{c}{EffDiff} \\
\cline{3-10}
\multicolumn{1}{c|}{} &\multicolumn{1}{c|}{}  &\multicolumn{2}{c|}{FID} & \multicolumn{2}{c|}{IS} & \multicolumn{2}{c|}{CLIP} & \multicolumn{1}{|c|}{WAN} 
& \multicolumn{1}{|c}{WAN$\textsuperscript{*}$}\\
\cline{3-10}
 &\multicolumn{1}{c|}{} &P$\downarrow$ & F$\downarrow$ & P$\uparrow$ & F$\uparrow$ & P$\uparrow$ & F$\downarrow$ & \multicolumn{1}{c|}{P$\uparrow$} & \multicolumn{1}{c}{F$\uparrow$}\\
\hline
\multirow{6}{*}{CelebA}
&Retrain &129.40 & 406.60 & 2.75 & 2.49 & 0.54 & 0.54 & 0.47 & -0.00 \\\cline{2-10}
&Max &403.00 & 532.60 & 2.68 & 2.63 & 0.44 & 0.66 & -0.22 & -0.61 \\
&Noisy &347.30 & 440.00 & 3.36 & 3.41 & 0.50 & 0.64 & 0.26 & -0.03 \\
&Retain &334.50 & 404.90 & 3.41 & 2.66 & 0.50 & 0.60 & 0.29 & -0.11 \\\cline{2-10}
&\textbf{\textit{SecureT2I}} & \textbf{166.80} & \textbf{406.00} & \textbf{2.29} & \textbf{2.81} & \textbf{0.54} & \textbf{0.56} & \textbf{0.29} & \textbf{0.06} \\
\hline
\multirow{6}{*}{Church}
&Retrain &109.00 & 462.10 & 2.61 & 2.43 & 0.58 & 0.55 & 0.36 & -0.05 \\\cline{2-10}
&Max &369.00 & 548.40 & 2.56 & 2.63 & 0.45 & 0.72 & -0.33 & -0.57 \\
&Noisy &227.80 & 452.70 & 3.28 & 3.10 & 0.51 & 0.72 & 0.33 & -0.02 \\
&Retain &178.80 & 461.40 & 2.94 & 2.89 & 0.55 & 0.63 & 0.32 & 0.04 \\\cline{2-10}
&\textbf{\textit{SecureT2I}} & \textbf{124.20} & \textbf{448.00} & \textbf{2.76} & \textbf{2.63} & \textbf{0.58} & \textbf{0.58} & \textbf{0.39} & \textbf{0.05} \\
\hline
\multirow{6}{*}{Bedroom}
&Retrain &111.00 & 462.30 & 3.04 & 3.04 & 0.69 & 0.55 & 0.54 & 0.11 \\\cline{2-10}
&Max &374.40 & 553.40 & 2.57 & 2.76 & 0.45 & 0.76 & -0.33 & -0.07 \\
&Noisy &336.60 & 514.40 & 3.34 & 3.54 & 0.50 & 0.69 & 0.12 & -0.12 \\
&Retain &355.50 & 518.80 & 3.24 & 3.38 & 0.50 & 0.69 & 0.04 & -0.20 \\\cline{2-10}
&\textbf{\textit{SecureT2I}} & \textbf{168.20} & \textbf{471.70} & \textbf{3.04} & \textbf{3.62} & \textbf{0.67} & \textbf{0.63} & \textbf{0.43} & \textbf{0.17} \\
\hline
\end{tabular}%
\label{tab:eff}
\end{table*}

\subsubsection{Evaluation of \textit{SecureT2I}.}
To address \textit{RQ2: Is there an effective approach that can balance performance on forbidden and permitted images?}, we propose \textit{SecureT2I}, which adopts a blurred version of the original image as the forbid set target to balance suppression and preservation, as detailed in Section~\ref{sec:def}. In our experiments, we use the blurred images as references for FID and CLIP calculations on the forbid set and apply the WAN$^*$ metric to better capture alignment with the vague target, enabling a more accurate assessment of each method’s effectiveness. We evaluate \textit{SecureT2I} across three mainstream text-guided image manipulation methods: DiffusionCLIP \cite{kim2022diffusionclip}, Asyrp \cite{kwon2022diffusion}, and EffDiff \cite{starodubcev2023towards}.
As shown in Tables \ref{tab:dclip}, \ref{tab:asyrp}, and \ref{tab:eff}, \textit{SecureT2I} achieves a significantly better balance between the permit and forbid sets compared to baseline methods. For example, on the CelebA dataset with DiffusionCLIP, while \textit{Retrain} achieves a WAN value of 0.67 on the permit set, \textit{SecureT2I} attains 0.44, substantially higher than unlearning methods such as \textit{Max} (-0.18), \textit{Noisy} (-0.30), and \textit{Retain} (-0.18). In terms of WAN$^*$, \textit{SecureT2I} scores 0.16, outperforming \textit{Retrain}’s 0.11, indicating improved suppression on the forbid set without sacrificing manipulation ability on permitted images. The FID scores further confirm this balance, with \textit{SecureT2I} recording 210.70 for the permit set and 413.40 for the forbid set, values close to those of \textit{Retrain} (135.00 permit, 401.30 forbid).
Similarly, on the Church dataset with Asyrp, \textit{SecureT2I} consistently surpasses unlearning-based methods in both WAN and WAN$^*$ metrics. The qualitative results in Fig. \ref{fig:example} further illustrate that \textit{SecureT2I} preserves image quality while effectively preventing unauthorized manipulations better than baseline methods.
Overall, these results demonstrate that \textit{SecureT2I} successfully balances performance on both forbidden and permitted images, providing a more effective and robust solution to secure image manipulation than existing approaches.

\begin{mdframed}
 {
    \textbf{Address RQ2:} We propose \textit{SecureT2I}, which uses a vague version of the original image as the target for the forbid set. Experiments on multiple datasets and models show that \textit{SecureT2I} outperforms existing methods by closely matching \textit{Retrain} on the permit set while achieving better results on the forbid set, offering a more balanced and effective solution than unlearning-based approaches.}
\end{mdframed}


\subsubsection{T-SNE Analysis.}

\begin{figure}
\centering
\subfigure[DiffusionCLIP]{\label{fig:tsne_dclip}
\includegraphics[width=0.31\textwidth]{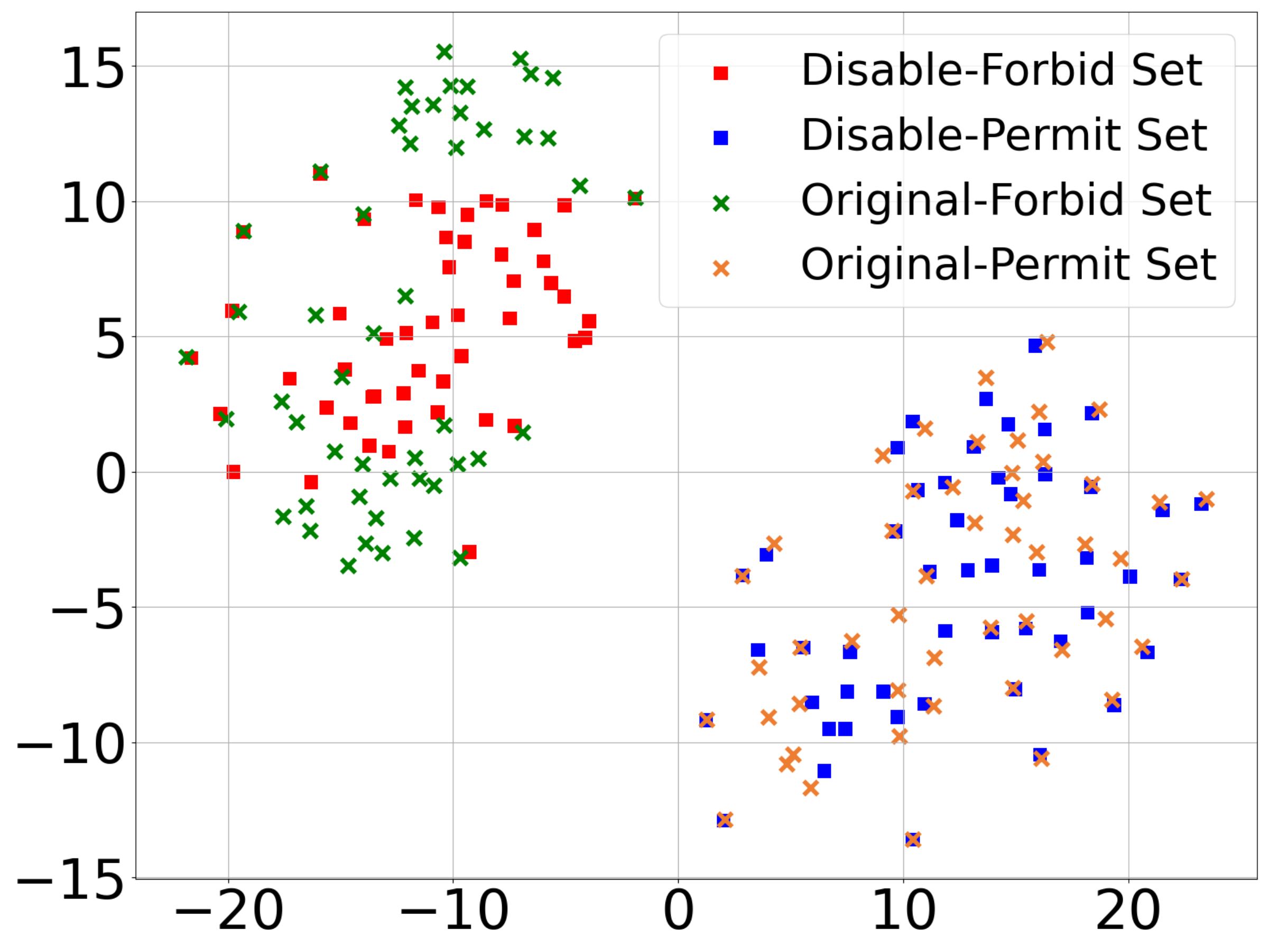}}
\subfigure[Asyrp]{\label{fig:tsne_asyrp}
\includegraphics[width=0.31\textwidth]{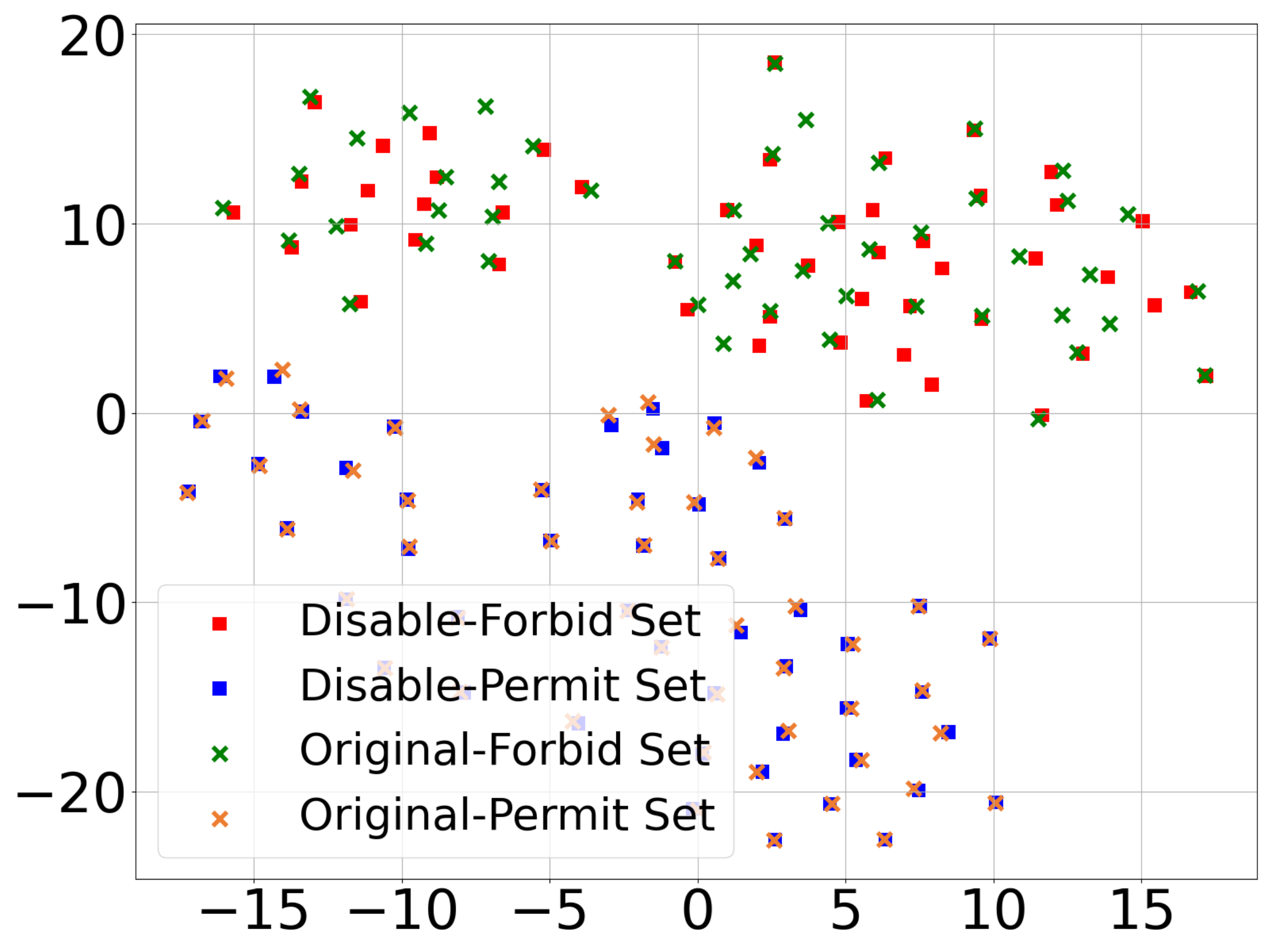}}
\subfigure[EffDiff]{\label{fig:tsne_eff}
\includegraphics[width=0.31\textwidth]{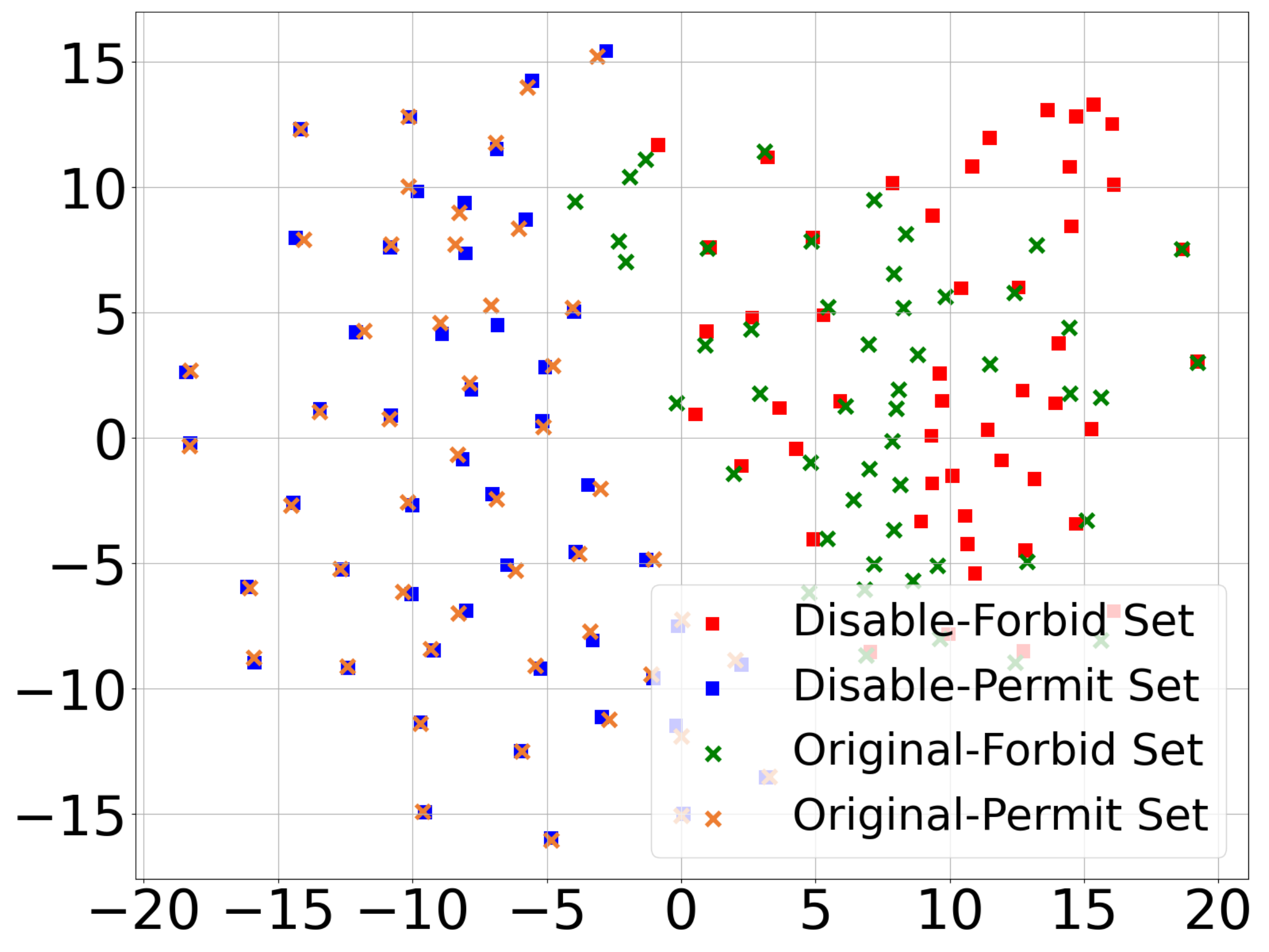}}
\subfigure[Max Loss]{\label{fig:tsne_max}
\includegraphics[width=0.31\textwidth]{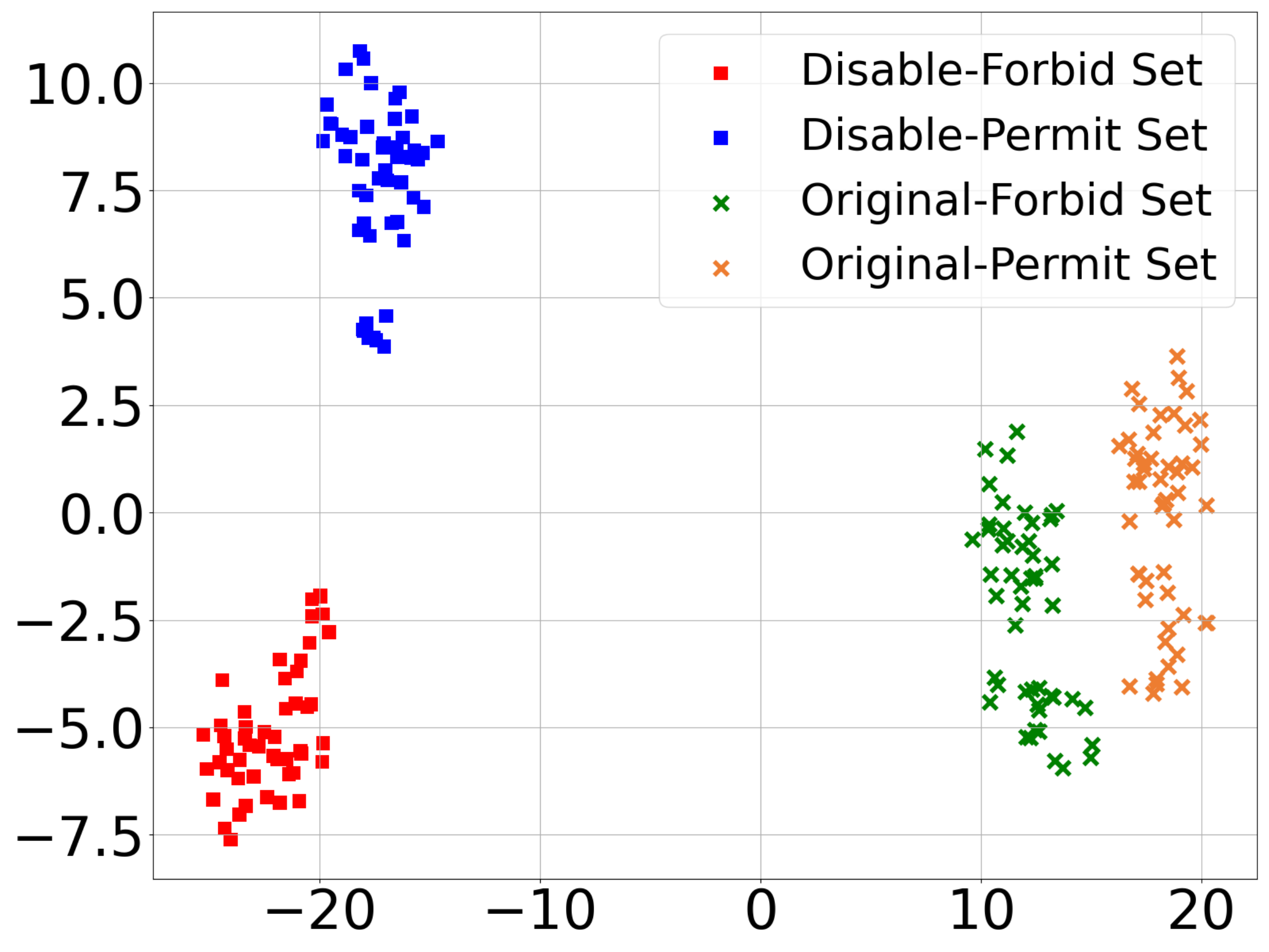}}
\subfigure[Noisy Label]{\label{fig:tsne_noisy}
\includegraphics[width=0.31\textwidth]{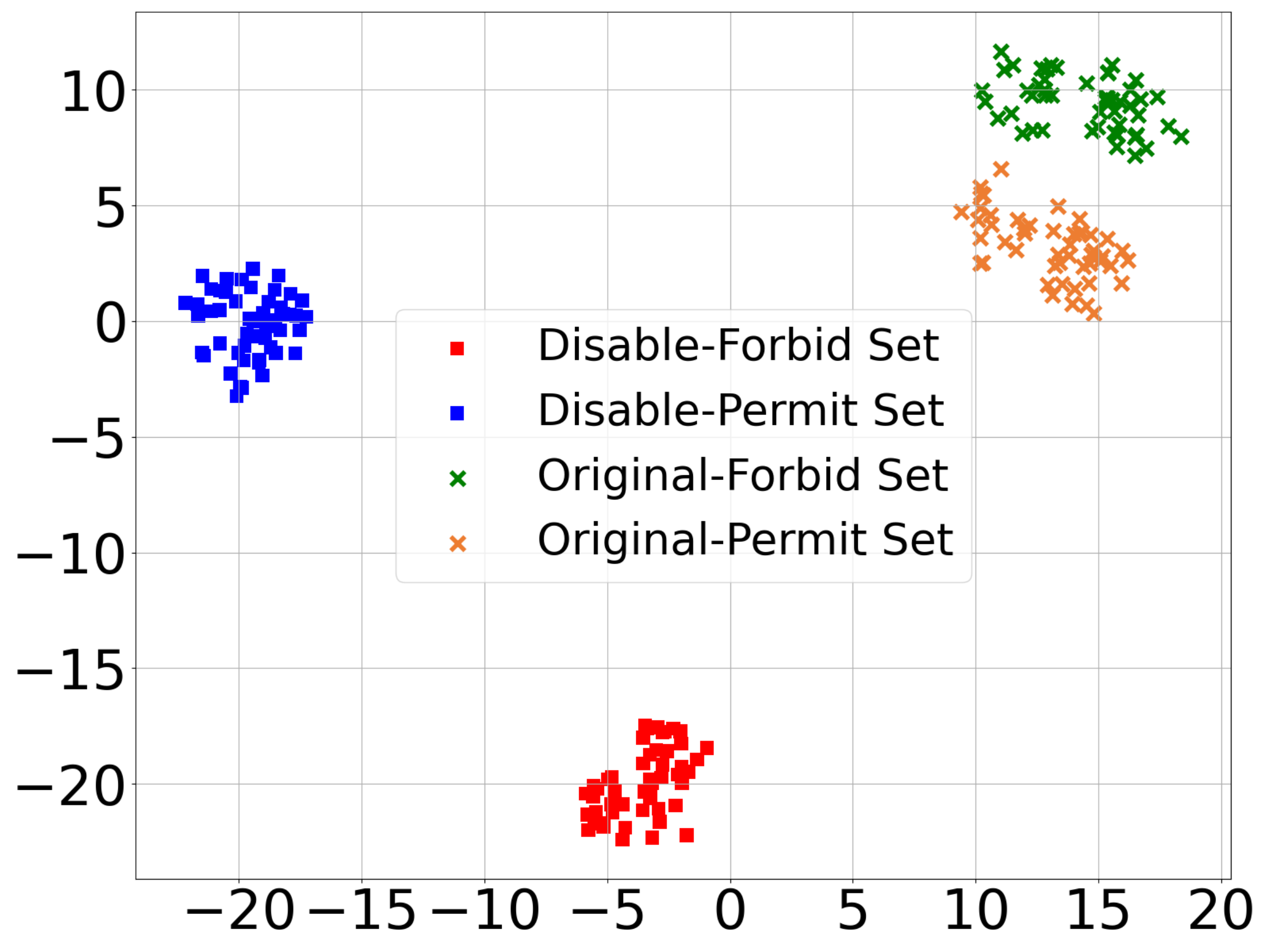}}
\subfigure[Retain Label]{\label{fig:tsne_retain}
\includegraphics[width=0.31\textwidth]{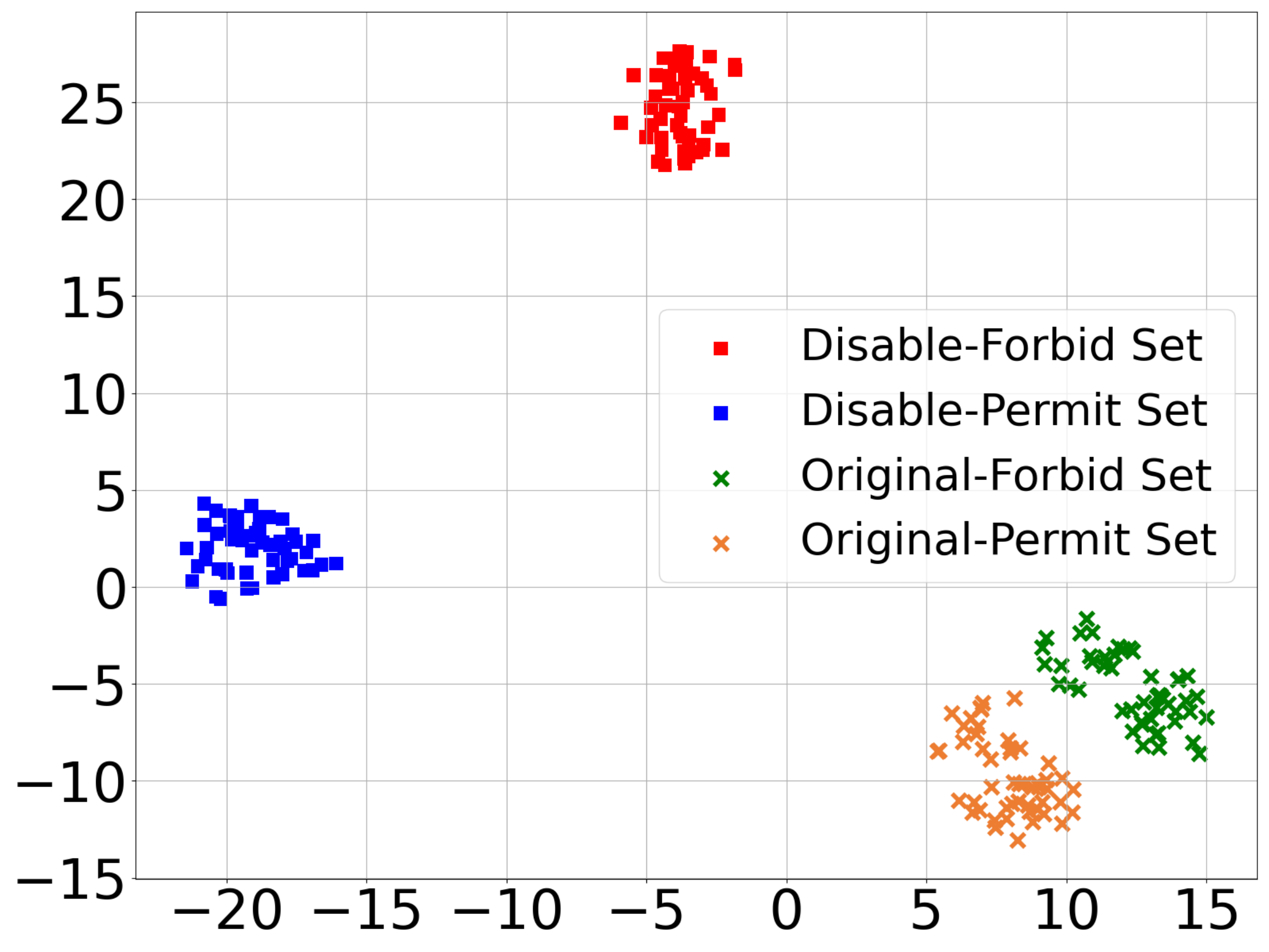}}
\caption{T-SNE illustration of the generated images from baseline methods, ground truth images and \textit{SecureT2I}.}
\label{fig:tsne}
\end{figure}

To provide a more comprehensive understanding of the superiority of our \textit{SecureT2I} over other baseline methods, we conduct an embedding visualization using the T-SNE technique. Specifically, for each of the three manipulation methods, we embed 50 images from both the forbid set and the permit set into a two-dimensional space, both before and after the prevention process. Additionally, the images generated by three baseline methods under DiffusionCLIP are also embedded for comparison.
As shown in Figs. \ref{fig:tsne_dclip}, \ref{fig:tsne_asyrp}, and \ref{fig:tsne_eff}, after applying the prevention mechanism, a clear separation emerges in the forbid set, where the generated images noticeably diverge from the originals. In contrast, in the permit set, the images remain closely clustered with their original counterparts. This demonstrates that \textit{SecureT2I} can effectively degrade manipulation performance on forbidden images while preserving generation quality for permitted images. Furthermore, the T-SNE visualizations for the three baselines tested on DiffusionCLIP, shown in Figs. \ref{fig:tsne_max}, \ref{fig:tsne_noisy}, and \ref{fig:tsne_retain}, reveal a substantial drift of embeddings for both the forbid and permit sets away from the original distributions. This indicates that these baselines significantly impair the model's overall performance, negatively impacting both forbidden and permitted images.

\subsection{Generalization to Unseen Images}

To evaluate the generalization ability of \textit{SecureT2I} to previously unseen data, we construct a held-out subset from both the permit and forbid sets by randomly sampling 10\% of images from each and excluding them entirely from the optimization process. These images are only used during evaluation to simulate real-world scenarios in which the system encounters inputs that were not part of the training distribution. Table~\ref{tab:unseen} presents the quantitative results across three datasets: CelebA, LSUN-Church, and LSUN-Bedroom. As the number of unseen images is relatively small, the Inception Score (IS) remains fixed at 1.00 for all methods. As a result, IS is excluded from normalization when computing the WAN and WAN$^\ast$ metrics.
From the results, we observe that \textit{SecureT2I} consistently outperforms the baselines in most settings. Specifically, it achieves lower FID and higher CLIP similarity on permit-set images, indicating high-quality and prompt-consistent edits. Meanwhile, the outputs for forbid-set images show reduced semantic similarity, as reflected by lower CLIP scores. These trends result in significantly higher WAN and WAN$^\ast$ values, suggesting that \textit{SecureT2I} effectively preserves authorized editing performance while suppressing unauthorized manipulations, even on inputs that were not seen during training. This confirms the method’s potential for generalization and its robustness to previously unseen samples.

\begin{table*}[htb]
\centering
\caption{Performance comparison of retraining, unlearning and \textit{SecureT2I} methods on unseen permit (P) and forbid (F) sets across three datasets using DiffusionCLIP.}
\small
\begin{tabular}{c|c|c|c|c|c|c|c|c|c}
\hline
\multirow{3}{*}{Datasets} &\multirow{3}{*}{Methods}  & \multicolumn{8}{c}{DiffusionCLIP} \\
\cline{3-10}
\multicolumn{1}{c|}{} &\multicolumn{1}{c|}{}  &\multicolumn{2}{c|}{FID} & \multicolumn{2}{c|}{IS} & \multicolumn{2}{c|}{CLIP} & \multicolumn{1}{|c|}{WAN} 
& \multicolumn{1}{|c}{WAN$\textsuperscript{*}$}\\
\cline{3-10}
 &\multicolumn{1}{c|}{} &P$\downarrow$ & F$\downarrow$ & P$\uparrow$ & F$\uparrow$ & P$\uparrow$ & F$\downarrow$ & \multicolumn{1}{c|}{P$\uparrow$} & \multicolumn{1}{c}{F$\uparrow$}\\
\hline
\multirow{6}{*}{CelebA}
&Retrain &180.30	&455.00	&1.00	&1.00	&0.73	&0.58 &0.21 &0.00  \\\cline{2-10}
&Max &412.02	&573.86	&1.00	&1.00	&0.52	&0.62 &-0.33  &-0.12  \\
&Noisy &418.58	&592.78	&1.00	&1.00	&0.53	&0.63 &-0.32  &-0.13 \\
&Retain &407.28	&564.83	&1.00	&1.00	&0.56	&0.66 &-0.27  &0.07  \\\cline{2-10}
&\textbf{\textit{SecureT2I}} & \textbf{128.69}	&\textbf{459.66}	&\textbf{1.00}	&\textbf{1.00}	&\textbf{0.78}	&\textbf{0.62} & \textbf{0.33} & \textbf{0.16} \\
\hline
\multirow{6}{*}{Church}
&Retrain &221.53	&499.78	&1.00	&.00	&0.80	&0.61 &0.29  &0.00  \\\cline{2-10}
&Max &389.57	&562.58	&1.00	&1.00	&0.62	&0.73 &-0.33 &0.06  \\
&Noisy &354.14	&566.39	&1.00	&1.00	&0.66	&0.69 &-0.20 &-0.07  \\
&Retain &332.25	&576.54	&1.00	&1.00	&0.67	&0.69 &-0.14  &-0.11 \\\cline{2-10}
&\textbf{\textit{SecureT2I}} & \textbf{197.68}	&\textbf{534.84} &\textbf{1.00}	&\textbf{1.00}	&\textbf{0.79}	&\textbf{0.70} & \textbf{0.31} & \textbf{0.10} \\
\hline
\multirow{6}{*}{Bedroom}
&Retrain &252.06	&549.83	&1.00	&1.00	&0.79	&0.70 &0.32 &0.00  \\\cline{2-10}
&Max &432.82	&633.86	&1.00	&1.00	&0.65	&0.76 &-0.23  &0.00  \\
&Noisy &405.53	&609.79	&1.00	&1.00	&0.62	&0.74 &-0.23  &-0.02  \\
&Retain &439.68	&610.56	&1.00	&1.00   &0.59	&0.72 &-0.33  &-0.13  \\\cline{2-10}
&\textbf{\textit{SecureT2I}} & \textbf{298.47}	&\textbf{581.87}	&\textbf{1.00}	&\textbf{1.00}	&\textbf{0.80}	&\textbf{0.74} & \textbf{0.25} & \textbf{0.10} \\
\hline
\end{tabular}%
\label{tab:unseen}
\end{table*}

\subsection{Impact of Different Vagueness Methods on Disabling Performance}

To address \textit{RQ3: What factors influence the effectiveness of SecureT2I?}, we investigate different ways to define the vague target for the forbid set, categorized into size-based compression (8×8, 16×16 \textit{SecureT2I}, 32×32) and filter-based blurring (Gaussian, Box, Motion). Size-based methods compress images and then resize them back to the original dimension, while filter-based methods apply various blurring techniques to introduce vagueness.
Our goal is to balance high-quality manipulation on the permit set and effective distortion on the forbid set. As shown in Table \ref{tab:vague}, \textit{SecureT2I} outperforms others by achieving the highest WAN and WAN$^*$ scores. Although Gaussian, Box, and Motion maintain decent permit set quality (WAN), they underperform on WAN$^*$, indicating insufficient forbidding effect. The 32×32 resizing shares similar limitations, while 8×8 overly distorts permitted images, harming quality despite yielding strong WAN$^*$ performance.
In summary, \textit{SecureT2I} best balances these trade-offs, delivering high-quality permitted outputs and effectively obscuring forbidden images, making it a reliable approach for secure text-guided image manipulation.

\begin{mdframed}
 {
    \textbf{Address RQ3:} The effectiveness of secure image manipulation depends heavily on the chosen vagueness method. Most resizing- or blurring-based approaches either insufficiently distort forbidden images or excessively degrade permitted ones. In contrast, \textit{SecureT2I} achieves the best balance with the highest WAN and WAN$^*$ scores, offering a robust and effective solution.
 }
\end{mdframed}

\section{Limitations}
\begin{table*}[htb]
\centering
\caption{Comparison of different vagueness methods on CelebA. P: Permit, F: Forbid. The row marked with `*' (\textbf{16x16*}) represents our \textit{SecureT2I} method.}
\small
\begin{tabular}{c|c|c|c|c|c|c|c|c}
\hline
\multirow{3}{*}{Vagueness} & \multicolumn{8}{c}{DiffusionCLIP} 
 \\
\cline{2-9}
 & \multicolumn{2}{c|}{FID} & \multicolumn{2}{c|}{IS} & \multicolumn{2}{c|}{CLIP} &\multicolumn{1}{|c|}{\multirow{1}{*}{WAN}} 
& \multicolumn{1}{|c}{\multirow{1}{*}{WAN$^*$}} \\
\cline{2-9}
 & P$\downarrow$ & F$\downarrow$ & P$\uparrow$ & F$\uparrow$ & P$\uparrow$ & F$\downarrow$ & \multicolumn{1}{c|}{P$\uparrow$} & \multicolumn{1}{c}{F$\uparrow$} \\
\hline
Retrain & 135.00 & 401.30 & 1.73 & 1.77 & 0.54 & 0.55 & 0.67 & 0.03 \\\hline
8x8 & 412.10 & 455.12 & 1.37 & 2.37 & 0.47 & 0.62 & -0.33 & 0.30 \\\hline
\textbf{16x16*} & \textbf{210.70} & \textbf{413.40} & \textbf{1.48} & \textbf{2.24} & \textbf{0.53} & \textbf{0.59} & \textbf{0.30} & \textbf{0.35} \\\hline
32x32 & 220.80 & 424.35 & 1.45 & 1.72 & 0.53 & 0.63 & 0.26 & 0.20 \\\hline
Gaussian & 215.70 & 423.73 & 1.43 & 1.71 & 0.53 & 0.60 & 0.25 & 0.08 \\\hline
Box & 228.30 & 429.34 & 1.48 & 1.75 & 0.53 & 0.59 & 0.29 & 0.02 \\\hline
Motion & 222.90 & 435.45 & 1.47 & 1.70 & 0.53 & 0.61 & 0.27 & 0.04 \\\hline
\end{tabular}%
\label{tab:vague}
\end{table*}
While \textit{SecureT2I} achieves effective suppression of unauthorized manipulations and preserves editing capabilities on authorized inputs, several limitations remain.
First, the current approach does not incorporate defenses against adversarial attacks on the diffusion model itself. Prior work has shown that diffusion models are vulnerable to small perturbations that can bypass editing restrictions \cite{truong2025attacks}, posing a threat to secure manipulation, especially in open-world scenarios where inputs may be intentionally tampered with. While our method focuses on semantic-level editing control via loss-based fine-tuning, it lacks robustness guarantees under such attacks. Incorporating adversarial training, robust noise prediction, or certified purification could enhance resilience and improve the model’s reliability in security-critical applications. Exploring these directions is an important avenue for future work.
Second, while our experiments cover a variety of prompts and image categories, we do not conduct a dedicated ablation study focusing on prompt variation. Robustness to diverse or rephrased prompts remains an open challenge in text-to-image systems, where different textual formulations with similar semantics may elicit inconsistent model behaviors. A systematic analysis of how prompt styles and manipulation intents influence suppression effectiveness would provide deeper insight into the stability of our method. We leave this investigation as an important direction for future work, especially for applications that require consistent control under natural prompt variability.
We view these limitations as important directions for future research, particularly in advancing secure text-guided manipulation systems that operate reliably across open-world distributions and under adversarial pressures.

\section{Conclusion}
In this paper, we introduced the critical and underexplored problem of secure image manipulation that prevents unauthorized edits based on text prompts. Our analysis revealed that existing solutions, such as retraining and unlearning, struggle to balance performance between authorized (permit) and unauthorized (forbid) images.
To address this challenge, we proposed \textit{SecureT2I}, a novel method that enables diffusion models to selectively disable unauthorized manipulations while preserving high-quality, legitimate edits. Experiments demonstrate that \textit{SecureT2I} significantly degrades manipulations on the forbid set and maintains fidelity on the permit set, outperforming all baselines. Notably, \textit{SecureT2I} also generalizes better to unseen images compared to existing methods.
Moreover, by systematically comparing various vagueness strategies for the forbid set, we found that resize-based vagueness offers the best trade-off between manipulation prevention and generation quality.
\textit{SecureT2I} sets a new benchmark for secure text-guided image manipulation and opens a promising direction toward more robust, scalable, and ethically aligned frameworks, marking a significant step in addressing ethical and copyright concerns in text-to-image generation.

%
%
%
\bibliographystyle{splncs04}
\bibliography{esorics25}

\section*{Appendix}
\addcontentsline{toc}{section}{Appendix}
\setcounter{section}{0}
\renewcommand{\thesection}{\Alph{section}}
\renewcommand{\thesection}{\Alph{section}}

\section{Implementation Details}
We develop our manipulation model based on a diffusion model with non-Markovian sampling, ensuring smooth transitions between input and output images during manipulation. The Adam optimizer is employed, with an initial learning rate of $8 \times 10^{-6}$. To balance suppression and retention, the weights $\lambda_{\text{forbid}}$ and $\lambda_{\text{permit}}$ are both set to $0.5$, enabling equal emphasis on prevention and retaining information. We randomly select 100 images from each of the three datasets. These images, paired with specific textual prompts (e.g., "beards" for CelebA-HQ), are used to fine-tune the manipulation model. This initial fine-tuning grants the model the ability to perform targeted manipulations based on the given prompt. After this, \textit{SecureT2I} is applied, further fine-tuning the model with the same set of $100$ images for an additional $15$ iterations. The model's performance is evaluated by averaging the results across five distinct manipulation features for each dataset.

\section{Dataset}
The details of the three datasets used in our evaluations are provided below:
\begin{itemize}
\item \textbf{CelebA-HQ} is a high-quality version of the CelebA dataset, consisting of 30,000 high-resolution images of celebrity faces. This dataset includes a rich variety of facial attributes such as gender, age, hairstyle, and facial expression, making it widely used in tasks like image generation, face editing, and attribute transfer.


\item \textbf{LSUN-Bedroom} is part of the Large-scale Scene Understanding (LSUN) dataset, containing millions of images from various scene categories. The Bedroom subset comprises around 3 million images of bedrooms, offering a rich source of visual diversity in terms of room layout, furniture arrangement, lighting, and style.

\item \textbf{LSUN-Church} is another subset of the LSUN dataset, featuring over 126,000 images of churches. These images include a wide range of church exteriors from different architectural styles and backgrounds, providing substantial diversity in terms of structure, weather conditions, and viewpoints.
\end{itemize}

\section{Evaluation Metrics}
\begin{itemize}
\item \textbf{FID} is a metric used to evaluate the quality of generated images by comparing the distribution of generated images to that of real images. It measures the similarity between features of images from a deep neural network (usually InceptionV3). Lower FID scores indicate that the generated images are closer to real images in terms of visual quality and diversity.

\item \textbf{IS} assesses the quality of generated images by evaluating two key factors: how diverse the generated images are and how well they represent a distinct object or scene. It uses a pre-trained inception network to classify images and calculates a score based on the entropy of the predicted labels. Higher IS scores indicate the images are both clear and varied.

\item \textbf{CLIP} embedding similarity evaluates the semantic correspondence between a generated image and a target image. CLIP projects both the generated and target images into a shared embedding space, where the similarity between their feature representations can be measured. A higher CLIP embedding similarity value indicates that the two images share similar semantic concepts and visual elements. This metric serves as an important indicator of how well the generated image captures the essence and meaning of the target image, helping assess the effectiveness of image generation or manipulation methods in terms of semantic consistency.
\end{itemize}

\end{document}